\pdfoutput=1
\documentclass[aps,prl,twocolumn,showpacs,reprint,superscriptaddress]{revtex4-2}

\usepackage{graphicx}
\usepackage{dcolumn}
\usepackage{amssymb}
\usepackage{bm}
\usepackage[utf8]{inputenc}
\usepackage{amsmath,amsfonts,amssymb}
\usepackage{color}
\usepackage[sf]{subfigure}
\usepackage{fouriernc}
\usepackage{xcolor}

\usepackage{etoolbox}

\usepackage{soul}

\usepackage[colorlinks=true,citecolor=blue,urlcolor=blue,linkcolor=blue,bookmarksopen]{hyperref}
\usepackage{comment}

\usepackage{amsmath,amssymb}
\usepackage{bm}
\usepackage{graphicx}

\renewcommand\d{\partial}

\renewcommand\d{\partial}
\newcommand\q{\mathbf{q}}
\newcommand\x{\mathbf{x}}

\newcommand\<{\langle}
\renewcommand\>{\rangle}

\begin{document}
	
\title{Multiple Magnetorotons and Spectral Sum Rules in Fractional Quantum Hall Systems}

\author{Dung Xuan Nguyen}
\affiliation{Brown Theoretical Physics Center and Department of Physics,
Brown University, 182 Hope Street, Providence, Rhode Island 02912,
USA}

\author{F.~D.~M.~Haldane}
\affiliation{Department of Physics, Princeton University, Princeton, New Jersey 08544, USA}

\author{E.~H.~Rezayi}
\affiliation{Physics Department, California State University Los Angeles, Los Angeles, California 90032, USA}

\author{Dam Thanh Son}
\affiliation{Kadanoff Center for Theoretical Physics, University of Chicago, Chicago, Illinois 60637, USA}

\author{Kun Yang}
\affiliation{NHMFL and Department of Physics, Florida State University, Tallahassee, Florida 32306, USA}
	
	\date{\today}
\begin{abstract}

We study numerically the the charge neutral excitations
(magnetorotons) in fractional quantum Hall systems, concentrating on
the two Jain states near quarter filling, $\nu=2/7$ and $\nu=2/9$, and
the $\nu=1/4$ Fermi-liquid state itself.  In contrast to the $\nu=1/3$
states and the Jain states near half filling, on each of the two Jain
states $\nu=2/7$ and $\nu=2/9$ the graviton spectral densities show
two, instead of one, magnetoroton peaks.  The magnetorotons have spin
2 and have opposite chiralities in the $\nu=2/7$ state and the same
chirality in the $\nu=2/9$ state.  We also provide a numerical
verification of a sum rule relating the guiding center spin $\bar s$
with the spectral densities of the stress tensor.

\end{abstract}

\maketitle

\textit{Introduction.}---The exploration of topological phases of
matter started with the discovery of the fractional quantum Hall (FQH)
effect \cite{Tsui,Laughlin:1983fy,Wen-topo}.  Under a strong magnetic
field, electrons in two dimensions form strongly correlated quantum
Hall systems. In the lowest Landau level (LLL) limit, the kinetic
energy of the electrons is constant, and the two-dimensional electron
system is driven to numerous exotic topological phases depending on
the filling fraction and the effective interactions. The topological
characteristics of the FQH ground state and its charged excitations
can be understood using the wavefunction approach, pioneered by
Laughlin~\cite{Laughlin:1983fy} and further developed by many other
authors~\cite{wf1,wf2,wf3,wf4,wf5}.  The idea of the composite fermion
(CF) by Jain \cite{Jain1,JainBook} provides an explanation of numerous
gapped FQH states observed in experiment \cite{FQHexp1,FQHexp2} and an
intuitive construction of model wavefunctions, and suggests the
abelian and non-abelian braiding statistic of quasi-particles and
quasi-holes \cite{wf6,Nayak}. Inspired by Jain's intuitive picture, a
field theory description of FQH states was developed (the
Halperin-Lee-Read, or HLR, theory), based on the idea of flux
attachment~\cite{FradkinLopez}, which predicts a gapless Fermi-liquid
state \cite{HLR} at half filling, which has been confirmed
experimentally \cite{CFExp}.

Recently, modifications to the HLR theory has been suggested to make
it consistent with the symmetries of a single Landau level.
Particle-hole symmetry, which has been long an issue~\cite{PH1,PH2},
is restored in the Dirac CF theory for FQH states near half
filling~\cite{Son:Dirac}.  The dipole coupling of the CF to the
electric field takes care of the consistency of the theory with
diffeomorphism~\cite{Nguyen:2017mcw,Du:2021pbc}.  In combination,
these two modifications lead to response functions consistent with all
known symmetries (see, e.g., Ref.~\cite{Hofmann:2021}).

An important feature of gapped quantum Hall states is the existence of
the neutral magnetoroton mode, first suggested by Girvin, MacDonald
and Platzman (GMP) \cite{GMP:1986} and later observed in
experiment \cite{Pinczuk1,Pinczuk2,Kukushkin}.  Though GMP originally
introduced the magnetoroton as a charge density wave, recent
works \cite{Haldane:2011,Haldane:2009,Golkar:2016thq}, employing the
lowest Landau level symmetries, suggest that magnetoroton has spin 2
and thus can be considered as a massive ``emergent graviton'' in FQH
systems.  The magnetoroton has been studied from many different
perspectives: by constructing the
wavefunction \cite{BoYang:wfroton,BoYang:wfroton1} (which conforms
with the spin-2 structure), as an excitation of the CF crossing
$\Lambda$ levels~\cite{Jain:rotonCF}, by exact
diagonalization~\cite{Jain:edroton,Jain:edroton1,Jain:edroton2,Jain:edroton3,Haldane:edroton,Moller:wfroton},
or within the Dirac CF
theory \cite{Golkar:2016thq,Nguyen:2017mcw,Nguyen:2018nematic}, where
it is interpreted as the shear deformation of the composite Fermi
surface.  In the latter studies, the chirality of the magnetoroton is
determined by the direction of the residual magnetic field seen by the
Dirac CFs, and the positions of the minima of the magnetoroton
dispersion relation match the experimental
results \cite{Pinczuk1,Pinczuk2,Kukushkin} rather well.

Very recently, the Dirac CF theory has been generalized to Jain states
near $\nu=\frac14$ \cite{Wang:DCF,Goldman:DCF,Nguyen:2021dirac}.  The
situation with the magnetoroton there seems to be very different from
that near $\nu=\frac12$: it is necessary~\cite{Nguyen:2021dirac} to
postulate a ``Haldane mode,'' i.e., an extra high-energy magnetoroton
(or multiple magnetorotons), in addition to the low-energy
magnetoroton that emerges from the dynamics of the CFs.  The
additional magnetoroton(s), which contribute to the projected static
structure factor (pSSF), are crucial for the Haldane
bound \cite{Haldane:bound}
\begin{equation}
	\label{eq:Haldanebound}
        S_4 \geq  \frac{|\bar s|}4 \,,
\end{equation}  
where $S_4$ is the coefficient of the leading $Q^4$~\footnote{Here we
define $Q=q \ell_B$ with $q$ being the momentum and $\ell_B$ the
magnetic length.} term in the pSSF, and $\bar s$ is the guiding center
spin.  On the LLL $\bar s = \frac12(\mathcal{S}-1)$ where
$\mathcal{S}$ is the Wen-Zee shift.  The extra magnetoroton was
heuristically suggested to arise from the microscopic structure of the
CF.  While the electric dipole moment of the CF is constrained by its
momentum \cite{Nguyen:2017mcw,Nguyen:2021dirac}, a higher moment
deformation of its shape could in principle generate a spin-2 mode,
which is the high-energy magnetoroton.

In this Letter we investigate the magnetoroton excitations, guided by
the FQH spectral sum rules \cite{Golkar,Nguyen2014LLL,
Nguyen:2021dirac} relating the chiral graviton spectral functions
with $\bar s$
and $S_4$.  These sum rules
constrain the spectral densities of the $\nu=p/(2np\pm1)$ states and
the Fermi-liquid-like state at $\nu=1/4$, and suggest the chiralities
of the magnetorotons there.  We calculate the spectral densities
numerically, from which we read out the chirality of the magnetorotons
and verify the sum rules.


\textit{Graviton spectral sum rules.}---In the LLL limit, when the
interacting energy is much smaller than the cyclotron energy, one can
obtain the exact sum rules involving the spectral densities of the
stress tensor \cite{Golkar,Nguyen2014LLL}.  In the complex coordinate
$z=x+iy$, the two components of the traceless part of the stress
tensor, $ T_{zz}=\frac{1}{4} \left( T_{xx}-T_{yy}-2i T_{xy}\right)$
and $T_{\bar{z}\bar{z}}=\frac{1}{4} \left( T_{xx}-T_{yy}+2i
T_{xy}\right)$, can be used to define two spectral
densities \cite{Golkar}
\begin{align}
	I_{-}(\omega) &= \frac{1}{N_e} \sum_n |\< n|\! \int\!d\x\, T_{zz} |0\>|^2 \delta(\omega-E_n), \label{eq:rhoT}
	\\
	I_{+}(\omega) &= \frac{1}{N_e} \sum_n |\< n| \! \int\!d\x\, T_{\bar z\bar z} |0\>|^2 \delta(\omega-E_n) \label{eq:rhoTb},
\end{align}
where $N_e$ is the total number of electrons, $|0\>$ is the ground
state, the sum is taken over all excited states $|n\>$ in the lowest
Landau level, and $E_n$ is the energy of the state $|n\>$ relative to
the ground state.  Physically, \eqref{eq:rhoT} and \eqref{eq:rhoTb}
are the densities of spin-2 states with opposite chiralities at
frequency $\omega$, and as such they depend on the microscopic details
of the FQH problem.  The expressions for the integrals of $T_{zz}$ and
$T_{\bar z\bar z}$ over space in terms of the LLL operators have been
derived in Ref.~\cite{Nguyen:2021Raman}.  We expect $I_-(\omega)$ and
$I_+(\omega)$ to vanish at frequencies below the energy gap; we also
expect them to rapidly go to 0 at energies much larger than the energy
scale set by the Coulomb interaction.  Using the $U(1)$ charge
conservation and the LLL limit of momentum conservation, one can
obtain the following exact sum rules \cite{Golkar, Nguyen2014LLL,
Nguyen:2021dirac} \footnote{The upper limit $\infty$ here means a
frequency which is much higher than the Coulomb gap but also smaller
the the cyclotron energy.}
\begin{align} \label{eq:sr1} \int_0^\infty \frac{d\omega}{\omega^2} \left[
I_-(\omega) + I_+(\omega) \right]& =
S_4,\\ \label{eq:sr2} \int_0^\infty \frac{d\omega}{\omega^2} \left[
I_-(\omega) - I_+(\omega) \right]& = \frac{\bar
  s}4 ~
 \left( = \frac{\mathcal{S}-1}{8} ~\text{on LLL}\right).
\end{align}
For derivations of the sum rules see
Refs.~\cite{Son:newtoncartan,Nguyen2014LLL, Nguyen:2021dirac}. Both
sum rules do not rely on a microscopic details and can be applied for
fractional quantum Hall states in the single Landau level limit, where
Landau-level mixing is ignored.  By definition $I_\pm(\omega)$ are
non-negative, therefore the sum rules imply the Haldane
bound \eqref{eq:Haldanebound}.  This bound is saturated if an only if
the FQH state is chiral, i.e., when one of the spectral densities
vanishes identically [i.e., $I_-(\omega)=0$ or $I_+(\omega)=0$].

\textit{General Jain states---} The Wen-Zee shift $\mathcal{S}$
of the general Jain state has been found previously \cite{Gromov:WZJain2,Nguyen:PH,Nguyen:2021dirac}
\begin{align}\label{eq:shiftplus}
		\nu_+=\frac{p}{2n p+1}, &\quad \mathcal{S}_+=p+2n,\\
		\label{eq:shiftminus}
		\nu_-=\frac{p}{2n p-1}, &\quad \mathcal{S}_-=-p+2n.
\end{align}  
The subscript index $+$ ($-$) corresponds to the residual magnetic
field seen by CFs being in the same (opposite) direction as the
applied magnetic field \cite{Nguyen:2017mcw,Nguyen:2021dirac}. The
direction of the residual magnetic field determines the chirality of
the low-energy magnetoroton, the one induced by the deformation of
composite Fermi surface.  If the residual magnetic field is in the
same (opposite) direction of external field, the low-energy
magnetoroton has negative (positive) chirality.  Consequently, we
expect a low energy peak in $I_-(\omega)$ ($I_+(\omega)$) of state
$\nu_+$ ($\nu_-$).

In Ref.~\cite{Nguyen:2021dirac} a Dirac CF model of the $\nu_\pm$
states were presented.  The model is supposed to be reliable at large
$p$ and its result for the spectral densities can be summarized as
\begin{equation}
		\label{eq:nuplus}
\begin{split}
  \nu_+: \quad & \frac{I_-(\omega)}{\omega^2} = \frac{p+1}8 \delta(\omega-\omega_L)
              + \frac{n-1}4 \delta(\omega-\omega_H),\\
            & I_+(\omega) = 0,  
\end{split}
\end{equation}
\begin{equation}
		\label{eq:numinus}
\begin{split}
  \nu_-: \quad & \frac{I_-(\omega)}{\omega^2} = \frac{n-1}4 \delta(\omega-\omega_H), \phantom{+\frac{0+0}8\delta(\omega-\omega_L)}\\
    & \frac{I_+(\omega)}{\omega^2} = \frac{p-1}8 \delta(\omega-\omega_L),
\end{split}    
\end{equation}
where $\omega_L$ and $\omega_H$ are the energies of the low- and
high-energy magnetorotons, respectively.  Note that the delta-function
$\delta(\omega-\omega_H)$ may be broadened by the decay of the
high-energy magnetoroton or splits into several peaks.  One can introduce
the integrated spectral densities
\begin{equation}
	\mathcal{I}_\mp =\int_0^\infty \frac{d\omega}{\omega^2} 
	I_\mp(\omega).
\end{equation}
The prediction of Ref.~\cite{Nguyen:2021dirac} reads
\begin{align}
	\label{eq:Iplus}
	\nu_+: \quad \mathcal{I}_-=\frac{p+2n-1}{8}, \quad \mathcal{I}_+=0,\\
	\label{eq:Iminus}
	\nu_-: \quad \mathcal{I}_-=\frac{n-1}{4}, \quad \mathcal{I}_+=\frac{p-1}{8}.
\end{align} 
and from the sum rule~(\ref{eq:sr1}) one finds the $S_4$ coefficient of the $\nu_\pm$ states:
\begin{align}\label{eq:S4plusminus}
   S_4(\nu_+) =\frac{p+2n-1}{8}\,, \quad
   S_4(\nu_-) =\frac{p+2n-3}{8}.
\end{align}  

Some remarks are in order. (i) For $n=1$ (near half filling), there is
no high-energy magnetoroton, and both $\nu_\pm$ states are
chiral. (ii) For $n\neq1$, only the $\nu_+$ state is chiral, with
$S_4$ saturating the Haldane bound, while the $\nu_-$ is not
chiral. (iii) Strictly speaking, the formulas presented above are
obtained in the large $p$ limit, so the application of these formulas
for the case of, say, $p=2$ should be taken with a grain of salt.  On
the other hand, one may expect that the qualitative statements about
the chirality of the magnetoroton modes are robust.

\textit{The Fermi-liquid states}.---In the
Fermi-liquid state with $\nu=1/2n$ the CFs are in zero emergent
magnetic field and form a Fermi liquid, whose excitations do not
contribute to the sum rule \eqref{eq:sr2}.  The only contribution to
the the sum rule \eqref{eq:sr2} is from the high-energy magnetoroton
(the Haldane mode).  Thus we find
\begin{equation}
	\label{eq:FL}
   \mathcal I_- -\mathcal I_+= \frac{n-1}4\,,
\end{equation}
Note that since the state is ungapped, the notion of $S_4$ does not
apply.  Naively we can associate $n-1$ with the guiding center spin
for of the Fermi-liquid state.  This, in turn, can be explained if one
thinks of the CF at $\nu=1/2n$ as a CF at $\nu=1/2$ state with
$2(n-1)$ flux quanta attached.  The CF at $\nu=1/2$ has no spin, while
each attached flux quanta increases the spin by $\frac12$.

Looking at Eqs. \eqref{eq:nuplus}, \eqref{eq:numinus},
and \eqref{eq:FL}, we notice that the contribution of the Haldane mode
to the guiding center spin
of states near $1/2n$ is universal.  In the Fermi-liquid state
$\nu=1/2$, the spectral densities $I_-(\omega)$ and $I_+(\omega)$
should be identical due to the particle-hole symmetry, therefore
$\mathcal I_- -\mathcal I_+=0$. (The same should be valid for the
PH-Pfaffian state~\cite{Son:Dirac}.)

\textit{Numerical results---}The graviton spectral function has been
investigated on both boson and fermion FQHE states which include
Moore-Read states~\cite{MOOREREAD} as well as Laughlin states in
Refs.~\cite{Liou:graviton-num, Haldane:graviton-num}. In this Letter we
present some results on the guiding center spin $\bar{s}$, the
coefficient $S_4$ of the $Q^4$ term in the pSSF,
and show the FQHE graviton spectral functions.

\begin{figure}[t]
\centering
{\includegraphics[height=2.2in,width=3.375in]{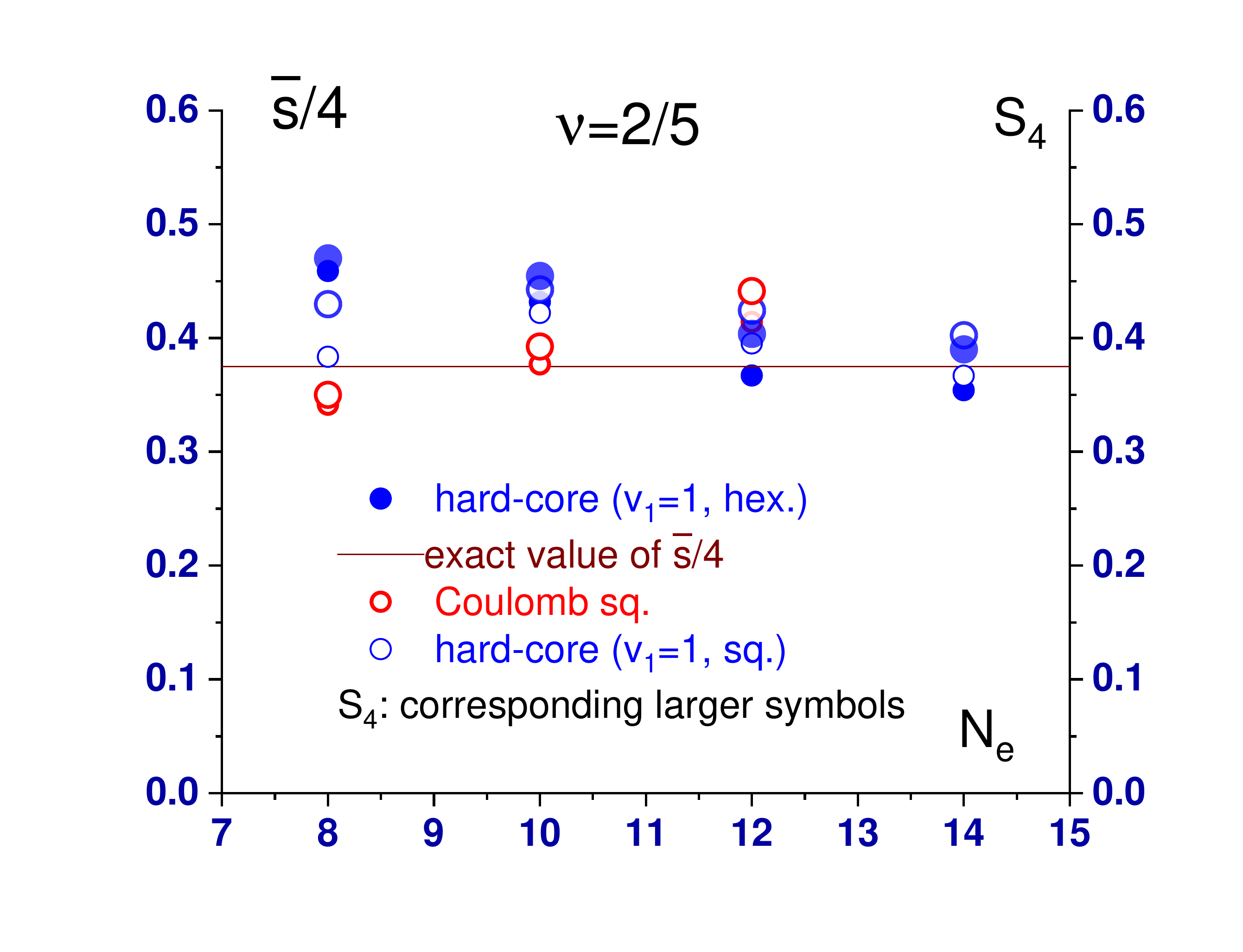}}
 \caption{\label{fig:SR-2-5}%
Graphical representation of $\bar{s}$ and $S_4$ obtained from the sum rules at electron filling factor of 2/5 for both the Coulomb and the hard-core potentials.  The latter is just a Haldane pseudo-potential with its value set to 1.
We see that $\bar{s}/4$ and  $S_4$ are mostly identical, which is expected for a chiral state. }
\end{figure}

We find good agreement with the theoretical predictions of
Eqs.~\eqref{eq:sr1} and \eqref{eq:sr2}.  We will present some of the
results in this Letter, delegating to the Supplement
Material some others not germane to the main topic of the Letter. To
obtain the correct sum rules for the Coulomb interactions it is
necessary to use the complete stress tensor given in the Supplement
Material
\cite{supp} to evaluate the spectral densities $I_-(\omega)$ and
$I_+(\omega)$.

\begin{figure}[t]
\centering
{\includegraphics[height=2.2in,width=3.375in]{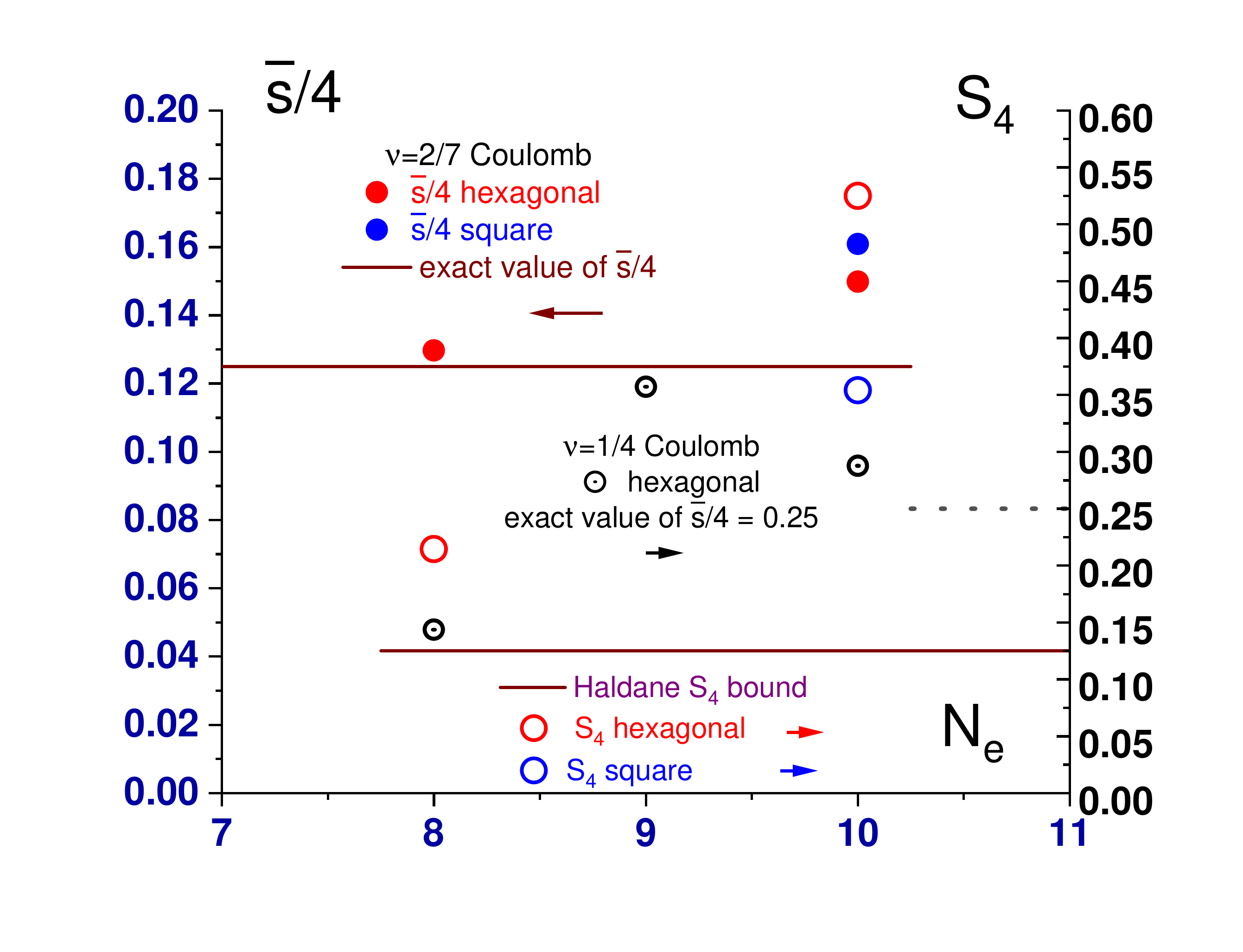}}
\caption{\label{fig:SR-2-7}%
Same as in Fig. \ref{fig:SR-2-5} but at electron filling factors  $\nu=2/7$ and $\nu=1/4$ for the Coulomb potential. The results for the 2/7 for $\bar{s}$ are given by the filled 
symbols and their values are on the left vertical axis (LVA). The RVA gives results for the $S_4$ of 2/7 (open symbols) and for $\bar{s}/4$ of $\nu=1/4$ (open symbols with a central dot).  The long lines show
the exact values of $\bar{s}/4$ on both vertical axes. The short dotted line gives the  same but 
for 1/4 state.
}
\end{figure}

In Figs.~\ref{fig:SR-2-5} and \ref{fig:SR-2-7} we present the results
for the sum rules for the Jain states at fillings 2/5, 2/7, and the
Fermi liquid state 1/4. We use the left vertical axis to represent
$\bar{s}/4$, while the right vertical axis represents $S_4$.  As
predicted by Haldane
\cite{Haldane:bound} only chiral states saturate the $S_4$
bound. Generic states such as the ones obtained from the Coulomb
interaction exceed this bound \cite{Golkar,Nguyen:2021dirac}. The
numerical results of $\bar{s}$ and $S_4$ of Jain states converge to
the theoretical predictions in Eqs. \eqref{eq:S4plusminus},
\eqref{eq:shiftplus} and \eqref{eq:shiftminus}.  We also numerically
verify the Wen-Zee shift of the Fermi liquid state $\nu=1/4$ predicted
by Eq.~\eqref{eq:FL}. The numerical results presented in
Figs.~\ref{fig:SR-2-5} and \ref{fig:SR-2-7} are highly non-trivial,
they are the first numerical check of the exact graviton spectral
sum rules in the LLL limit.

\begin{figure}[ht]
\centering
{\includegraphics[height=2.2in,width=3.375in]{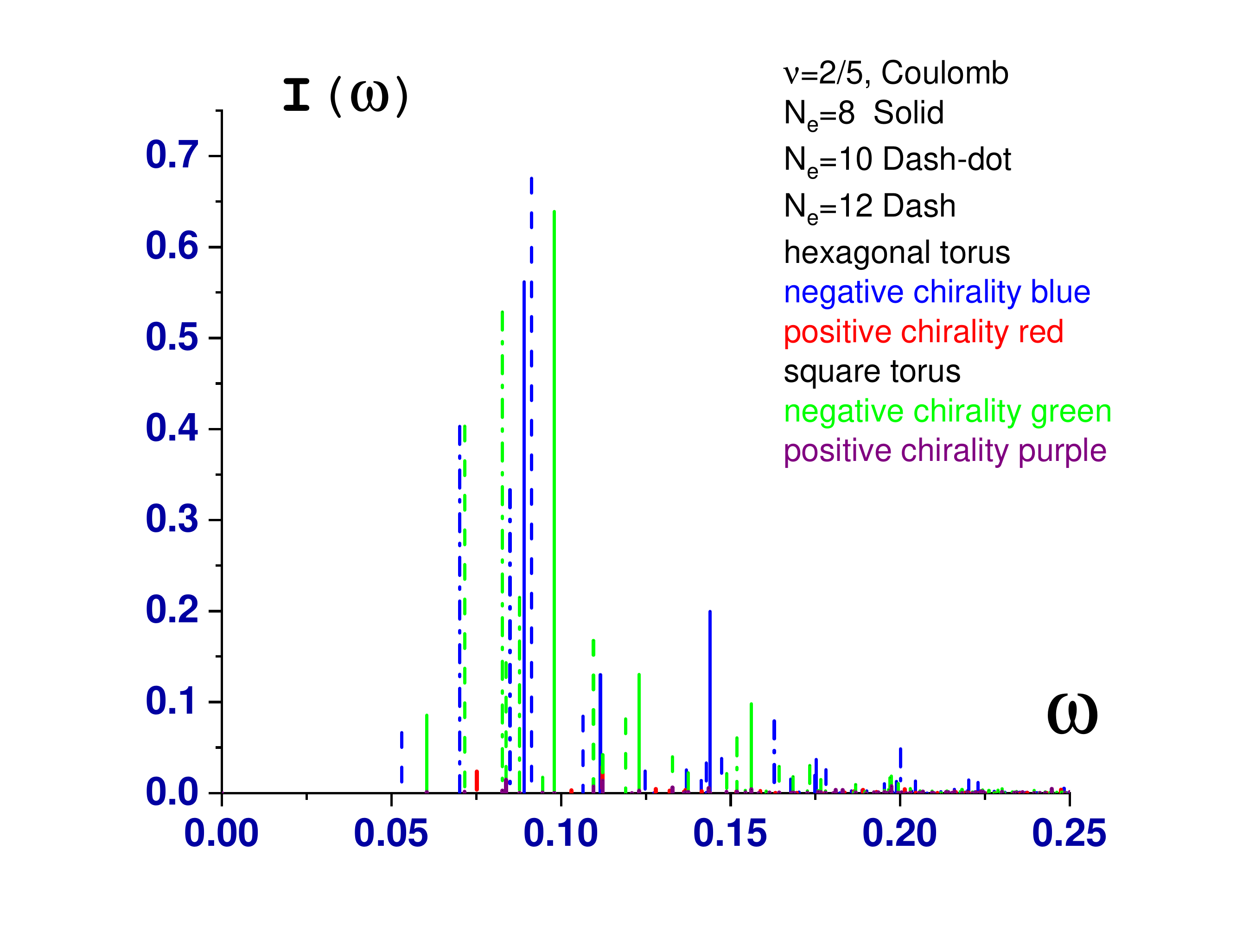}}
\caption{\label{fig:GS-2-5}%
The graviton spectral function for the filling factor $\nu=2/5$ 
and pure Coulomb potential. The positive chirality total strength of the spectral function
is much less than the negative chirality.  As a result we have normalized the spectrum 
of both chiralities by the total weight of the gravitons with negative chirality. 
The quasi-chiral nature can be attributed
to dearth of quasi-holes of the chiral parent, which is 1/3 for the hierarchy~\cite{Haldane_83_hierarchy}. Similarly, in Jain's CF approach, the spectrum is nearly chiral if  
there is not any opposing flux (to the attached ones) from the filled LLs.
}
\end{figure}

\begin{figure}[ht]
\centering
{\includegraphics[height=2.2in,width=3.375in]{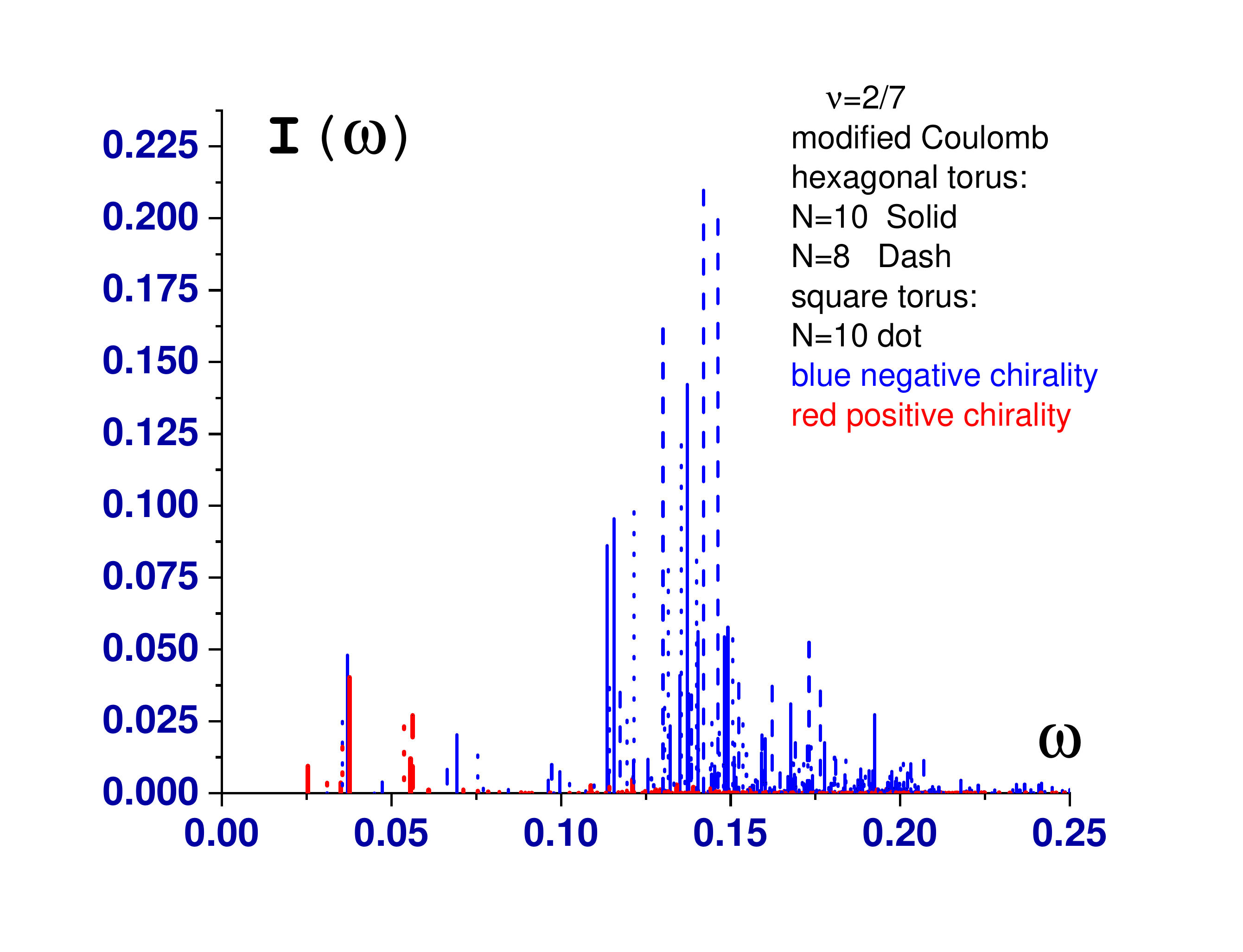}}
\caption{\label{fig:GS-2-7}%
Same as in Fig. \ref{fig:GS-2-5} but for $\nu=2/7$ filling and with the modified 
Coulomb.  A clear separation of predominantly positive chirality is seen on the left. On the right is the spectrum for the negative chirality, which appears to contain the main weight of total spectral density. Here we have the opposite case from 2/5.  Hierarchy involves quasiholes and filled LLs in the Jain construction have opposite B-field to the attached fluxes.
}
\end{figure}

\begin{figure}[t]
\centering
{\includegraphics[height=2.2in,width=3.375in]{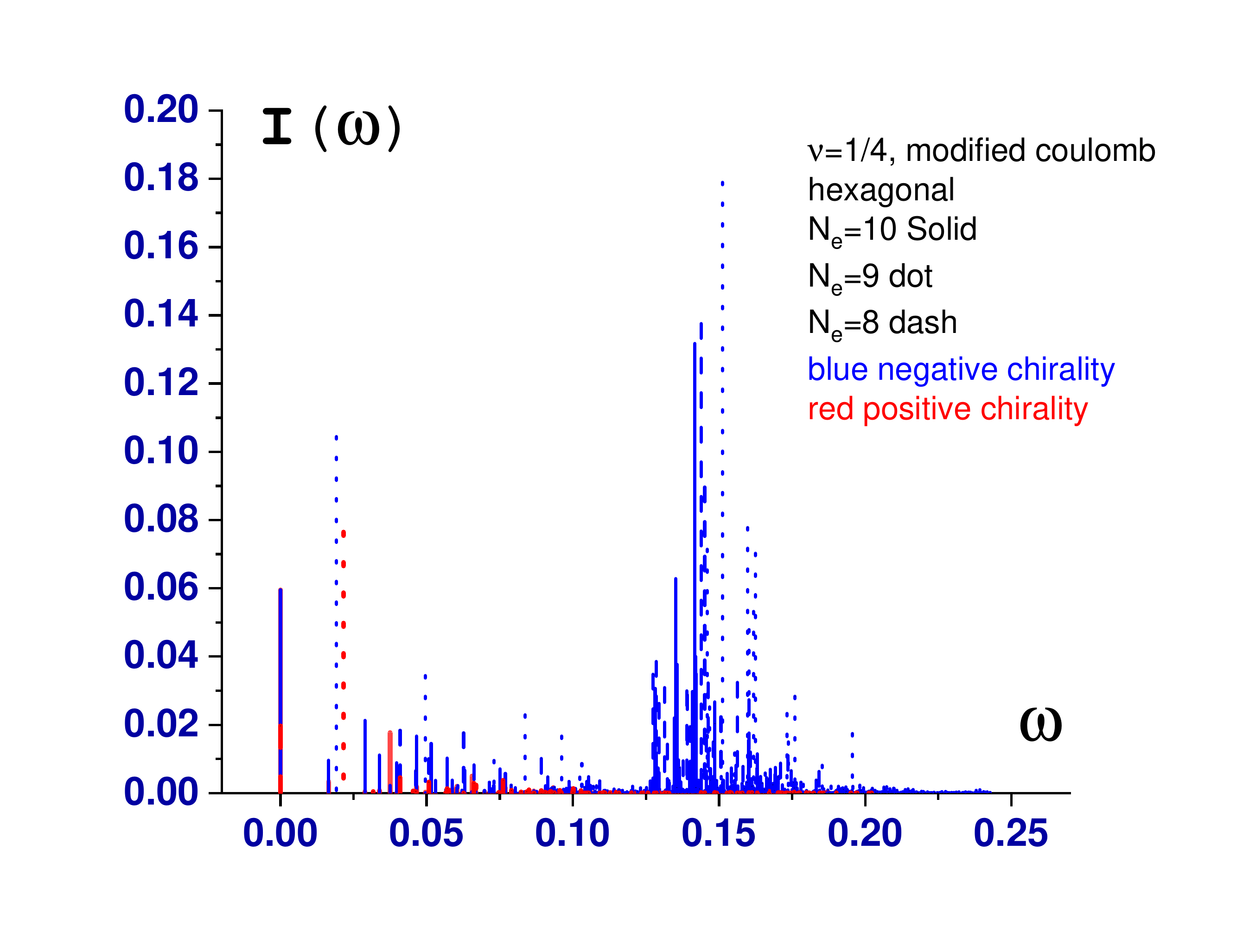}}
\caption{\label{fig:GS-1-4}%
The spectral function but for $\nu=1/4$ composite Fermi liquid for the modified Coulomb potential.  Note that the low energy spectrum has an equal weight for both chiralities with our normalization. This means the contribution to $\bar{s}$ is solely due to the Haldane mode. 
}
\end{figure}

\begin{figure}[t]
\centering
{\includegraphics[height=2.2in,width=3.375in]{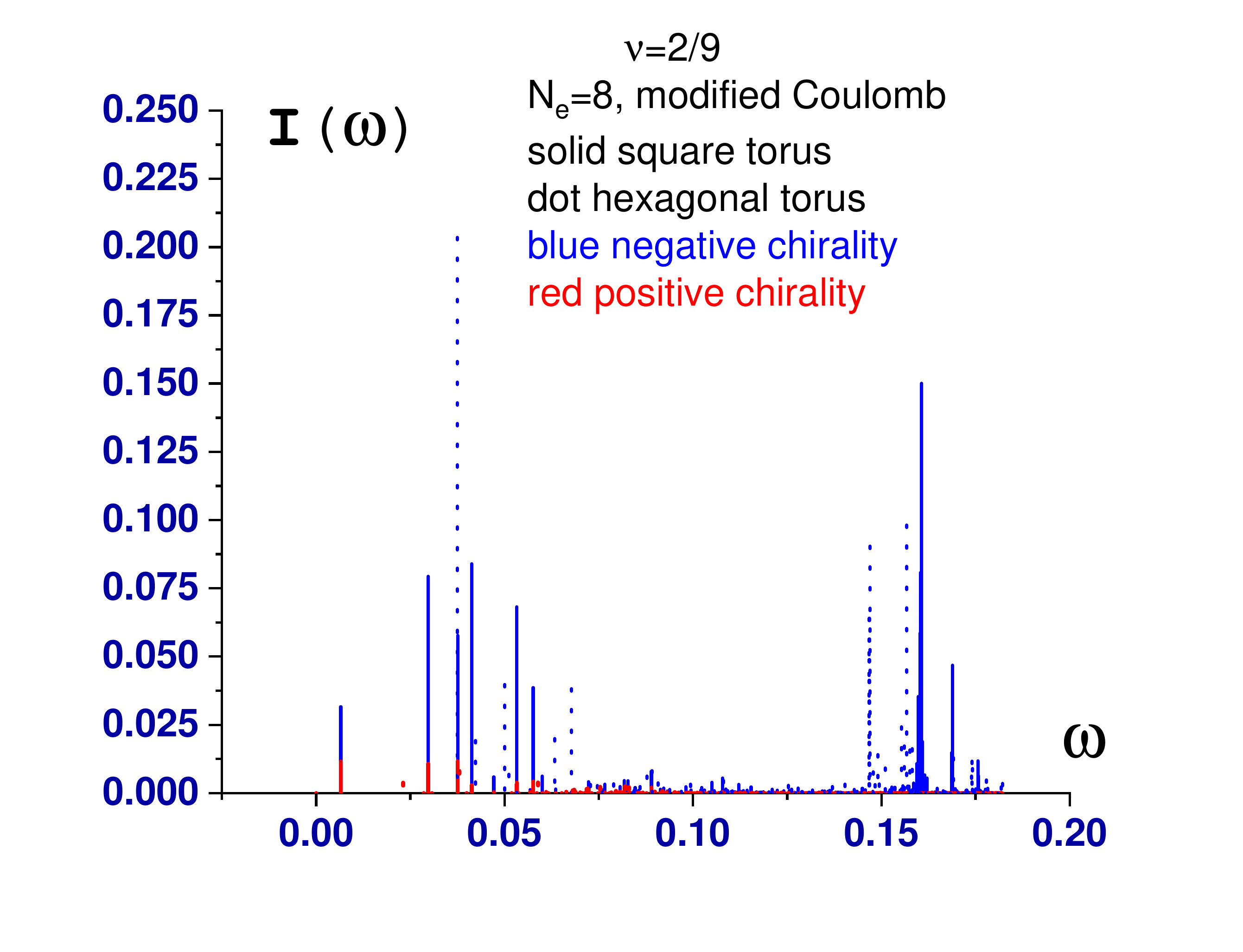}}
\caption{\label{fig:GS-2-9}%
The spectral function but for $\nu=2/9$ with the modified Coulomb potential. The spectrum is nearly completely chiral as would be expected since it is to the right of the 1/5 in
the hierarchy or involves filled LLs with no opposing flux in the Jain's construction as in 2/5  filling.  Two well separated sectors with
negative chirality can be seen clearly. Here the size is too small for us to get a meaningful
$\bar{s}$ and $S_4$.  However, the qualitative feature of having two separated regions of
nontrivial graviton spectral density will persist for larger sizes.
}
\end{figure}

Figures \ref{fig:GS-2-5}--\ref{fig:GS-2-9} present the FQH graviton
spectral functions for $\nu=2/5$, $2/7$, $1/4$, and $2/9$.  For
$\nu=2/5$, we use just the Coulomb potential since we want to compare
it to the spectrum obtained with the hard-core potential, but for the
remaining three fractions we calculate the spectral functions using a
modified version of Coulomb.  The two forms give the same qualitative
results for the distributions of the graviton weights, particularly in
instances where there are two distinct energy sectors.

All the theoretical predictions on graviton's chirality for the
general Jain states from Dirac CF model of
Ref.~\cite{Nguyen:2021dirac} are confirmed numerically.  For $n=1$,
the chirality of gravitons are determined by the residue magnetic
field seen by the CFs, therefore the graviton of $\nu=2/5$ has
negative chirality as showed in Fig.~\ref{fig:GS-2-5}. With $n=2$ the
chirality of low energy graviton is also determined by the residue
magnetic field, and the chirality of the high energy graviton is
universal for all Jain states near $1/4$. The predictions are
confirmed in Figs.~\ref{fig:GS-2-7} and \ref{fig:GS-2-9}.

Interestingly, the graviton spectral functions of Fermi-liquid state
$\nu=1/4$ in Fig.~\ref{fig:GS-1-4} shows the high energy graviton with
expected chirality. From both the shift sum rule and the spectral
densities, we see that the Haldane mode is universal for all FQH
states near $1/4$: it does not care if there is a composite Fermi
surface or if there is a residue magnetic field.
Fig.~\ref{fig:GS-1-4} also shows the low-energy excitations of both
chirality with equal weight.  We expect that the spectral densities of
the Fermi-liquid state $\nu=1/2$ share the same feature with the low
energy spectral densities of $\nu=1/4$ with $I_-(\omega)$ and
$I_+(\omega)$ being similar, the only difference is the non-appearance
of the Haldane mode.

\textit{Conclusion.}---In this Letter, we compute numerically the
graviton spectral densities for the $\nu=2/7$, $2/9$, and $1/4$
states.  For the first two states, we observe in the spectral
densities two magnetoroton peaks, one at low energy with chirality
that depends on the residual magnetic field acting on the CFs (and
thus are opposite for the $\nu=2/7$ and $2/9$ states), and one at high
energy with the same chirality for the two states.  The higher-energy
magnetoroton is also observed in a spectral density of the $\nu=1/4$
state, and this magnetoroton has approximately the same energy in all
three filling fractions.  The result is consistent with the
two-magnetoroton model of FQH states near $1/4$ proposed in
Ref.~\cite{Nguyen:2021dirac}. In addition, we have verified the FQH
graviton spectral sum rules for the two Jain states and the Fermi
liquid state $\nu=1/4$.

We hope that our results will motivate the experimental exploration of
the magnetoroton spectrum of the FQH states near $\nu=1/4$, as well as
more detail study of the Haldane mode which may reveal its nature.


This work is supported, in part, by the US Department of Energy, Basic Energy Sciences grant \protect{DE-SC0002140} (FDMH, EHR, KY), U.S. DOE grant
No.\ DE-FG02-13ER41958 (DTS),  a Simons Investigator grant and by the Simons
Collaboration on Ultra-Quantum Matter, which is a grant from the
Simons Foundation (651440, DTS). KY’s work was performed at the National High Magnetic
Field Laboratory, which is supported by National Science
Foundation Cooperative Agreement No. DMR-1644779, and
the State of Florida. DXN was supported by Brown
Theoretical Physics Center.

\textit{Note:} While this manuscript is being prepared, we became
aware of the work \cite{Gromov-Papic-Balram} that also discusses the
extra graviton mode in the FQH states $\nu=2/7$ and $\nu=2/9$.  We thank
Ajit C.\ Balram, Zhao Liu, Andrey Gromov, and Zlatko Papi\'c for
sharing their manuscript with us before publication.

\bibliography{GS}
\clearpage
\onecolumngrid

\begin{center}
	\textbf{\large --- Supplementary Material ---\\ Multiple Magnetorotons  and Spectral Sum Rules in Fractional Quantum Hall Systems}\\
	\medskip
	\text{ Dung Xuan Nguyen, E. H. Rezayi,  Dam Thanh Son, and Kun Yang }
\end{center}
\setcounter{equation}{0}
\setcounter{figure}{0}
\setcounter{table}{0}
\setcounter{page}{1}
\makeatletter
\renewcommand{\thesection}{S\arabic{section}}
\renewcommand{\theequation}{S\arabic{equation}}
\renewcommand{\thefigure}{S\arabic{figure}}
\renewcommand{\bibnumfmt}[1]{[S#1]}

\section{Stress tensor operator on a general Landau level}\label{sec:Tfull}
In order to verify numerically the spectral sum rules, we need to use
the single Landau level limit of the holomorphic components of the
stress-tensor operator. The zero momentum stress tensor projected to a
general Landau level $N$ is given in Ref.~\cite{Nguyen:2021Raman},
\begin{equation}\label{eq:Tfull}
  \int\!d\x\, T_{ij}(\x) =\frac12 \sum_\q
  \frac{q_i q_j}q \frac\d{\d q} \left\{
  e^{-\frac{q^2 \ell_B^2}{2}} \left[L_N\left(\frac{q^2 \ell_B^2}{2}\right)\right]^2 V(q) \right\}
  \bar\rho(\q) \bar\rho(-\q) .
\end{equation}
with $L_N(x)$ being the Laguerre polynomial and $V(q)$ the Fourier transformed of the two-body interaction,
	\begin{equation}
		V(\mathbf{q})=\int d^2 \x \, V(\x) e^{i \q \cdot \x} .
	\end{equation} 
Since we assume the interaction is rotationally invariant, we expect that $V(\mathbf{q})$ only depends on 
\begin{equation}
	q=\sqrt{\q \cdot \q}.
\end{equation}
In the complex coordinate, we define the complex momentum as 
\begin{equation}
	q_z=\frac{1}{2}(q_x+iq_y), \quad q_{\bar{z}}=\frac{1}{2}\left(q_x-iq_y	\right).
\end{equation}
Using Eq.~\eqref{eq:Tfull}, we can write the holomorphic components of
the stress tensor at zero momentum as
\begin{equation}\label{eq:Tzz}
   \int\!d\x\, T_{zz}(\x) = \frac12 \sum_\q
  \frac{q_z q_z}q \frac\d{\d q} \left\{
  e^{-\frac{q^2 \ell_B^2}{2}} \left[L_N\left(\frac{q^2 \ell_B^2}{2}\right)\right]^2 V(q) \right\}
  \bar\rho(\q) \bar\rho(-\q) ,
\end{equation}	
\begin{equation}\label{eq:Tzbzb}
  \int\!d\x\, T_{\bar{z}\bar{z}}(\x) =\frac12 \sum_\q
  \frac{q_{\bar z } q_{\bar z}}q \frac\d{\d q} \left\{
  e^{-\frac{q^2 \ell_B^2}{2}} \left[L_N\left(\frac{q^2 \ell_B^2}{2}\right)\right]^2 V(q) \right\}
  \bar\rho(\q) \bar\rho(-\q) .
	\end{equation}
In numerical calculations, we use the explicit form of the stress
tensor in Eqs.~\eqref{eq:Tzz}, \eqref{eq:Tzbzb} and the ground state
from exact diagonalization to obtain the spectral functions
$I_-(\omega)$, $I_+(\omega)$.

\section{Numerical procedure}

Our numerical studies are based on the evaluation of two stress tensor
spectral densities
\begin{align}
	I_{-}(\omega) &= \frac{1}{N_e} \sum_n |\< n|\! \int\!d\x\, T_{zz} |0\>|^2 \delta(\omega-E_n), 
	\\
	I_{+}(\omega) &= \frac{1}{N_e} \sum_n |\< n| \! \int\!d\x\, T_{\bar z\bar z} |0\>|^2 \delta(\omega-E_n) ,
\end{align}
where the explicit expression of the full stress tensor at zero
momentum is given in Eqs.~\eqref{eq:Tzz} and \eqref{eq:Tzbzb}. Note
that the expressions of the stress-tensor components~\eqref{eq:Tzz}
and \eqref{eq:Tzbzb} differs from the ones used in
Refs.~\cite{Liou:graviton-num,Haldane:graviton-num,BoYang:stress,Liu:stress}. As
explained in Ref.~\cite{Nguyen:2021Raman}, the expressions in
Ref.~\cite{Liou:graviton-num,Haldane:graviton-num,BoYang:stress,Liu:stress}
includes only the ``kinetic part'' but omits the ``potential part'' of
the stress tensor.

With the short-range pseudo-potential, the two expressions for the
stress tensor give the same numerical result, since one can show that
the contribution to zero momentum stress tensor from interaction
annihilates the ground state~\cite{Nguyen2014LLL}. However, to obtain
the correct sum rules \eqref{eq:sr1}, \eqref{eq:sr2} for the Coulomb
interactions, it is necessary to use the complete stress tensor
\eqref{eq:Tzz} and \eqref{eq:Tzbzb} to evaluate the spectral densities
$I_-(\omega)$ and $I_+(\omega)$. The end result is that the LLL
Coulomb potential is modified by adding to $1/Q$ its
cube. The first Haldane pseudo-potential is unaffected for both bosons and fermions.  We've performed numerical calculation of both versions in this work.

The computer resources required for these types of calculations are
much more demanding than those in obtaining the Hall viscosity
\cite{Read:Hallviscos} from the Berry curvature. The gain is the
calculation of $S_4$ and the spectral functions of Jain (or hierarchy)states, which are new. In this Letter and in the Supplement Material,
we present the numerical results of the spectral densities
$I_-(\omega)$ and $I_+(\omega)$ as well as the integrated spectral
densities $\mathcal{I}_-$ and $\mathcal{I}_+$.\vspace{-0.5cm} 

\section{Supplemental Figures}
\begin{figure}[ht]
	\hfill
	\subfigure[$\nu=1/3$]{\includegraphics[width=8cm]{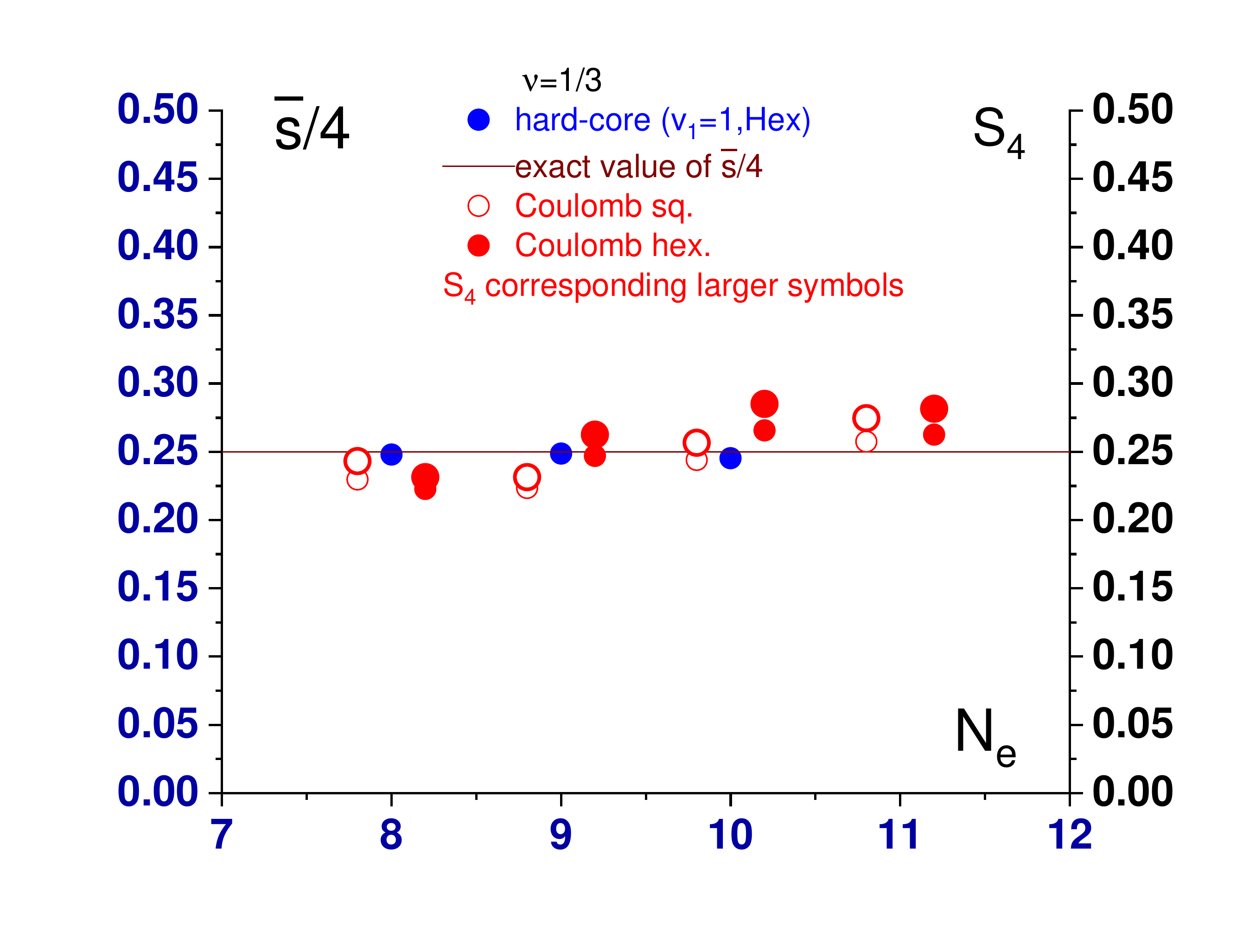}}
	\hfill
	\subfigure[$\nu=2/3$]{\includegraphics[width=8cm]{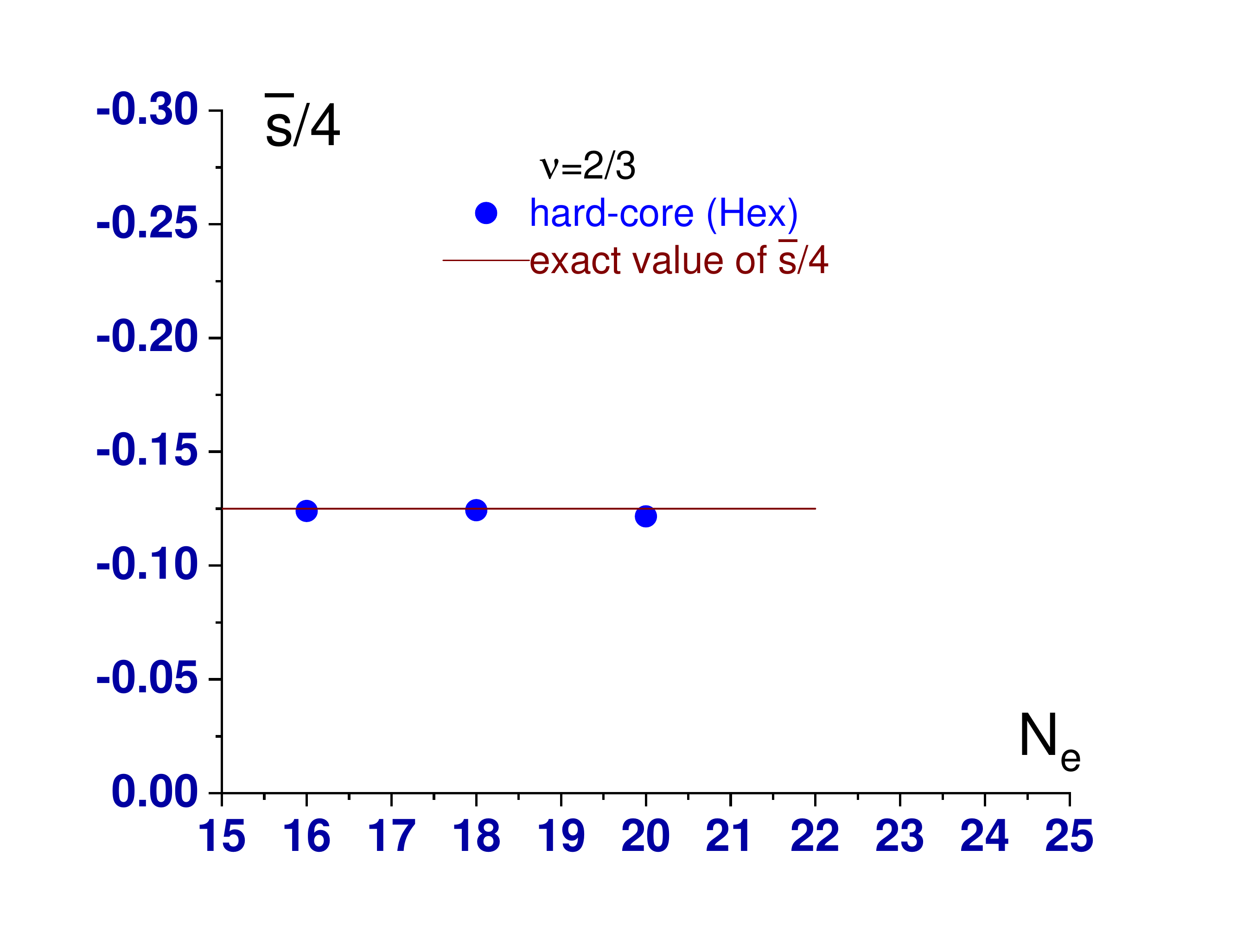}}
	\hfill
	\caption{%
(a) Graphical representation of $\bar{s}$ and $S_4$ obtained from the sum rules at electron filling factor of $\nu=1/3$ for both the Coulomb and the hard-core potentials. We see that $\bar{s}/4$ and  $S_4$ are mostly identical, which is expected for a chiral state. (b) Graphical representation of $\bar{s}$ obtained from the sum rules at electron filling factor of $\nu=2/3$ for hard-core potentials. }
\label{fig:supplemental1}        
\end{figure}

In this supplemental section, we show some more numerical results of the shift sum rule for  $\nu=1/3$ and $\nu=2/3$ in Fig.~\ref{fig:supplemental1}.
The chiralities of graviton for these states come out as expected with negative for  $\nu=1/3$  and positive for  $\nu=2/3$.  	We see that the sum rules are again confirmed by numerical calculations.\vspace{-0.3cm} 
%

\begin{figure}[ht]
	\centering
	{\includegraphics[height=2.2in,width=3.375in]{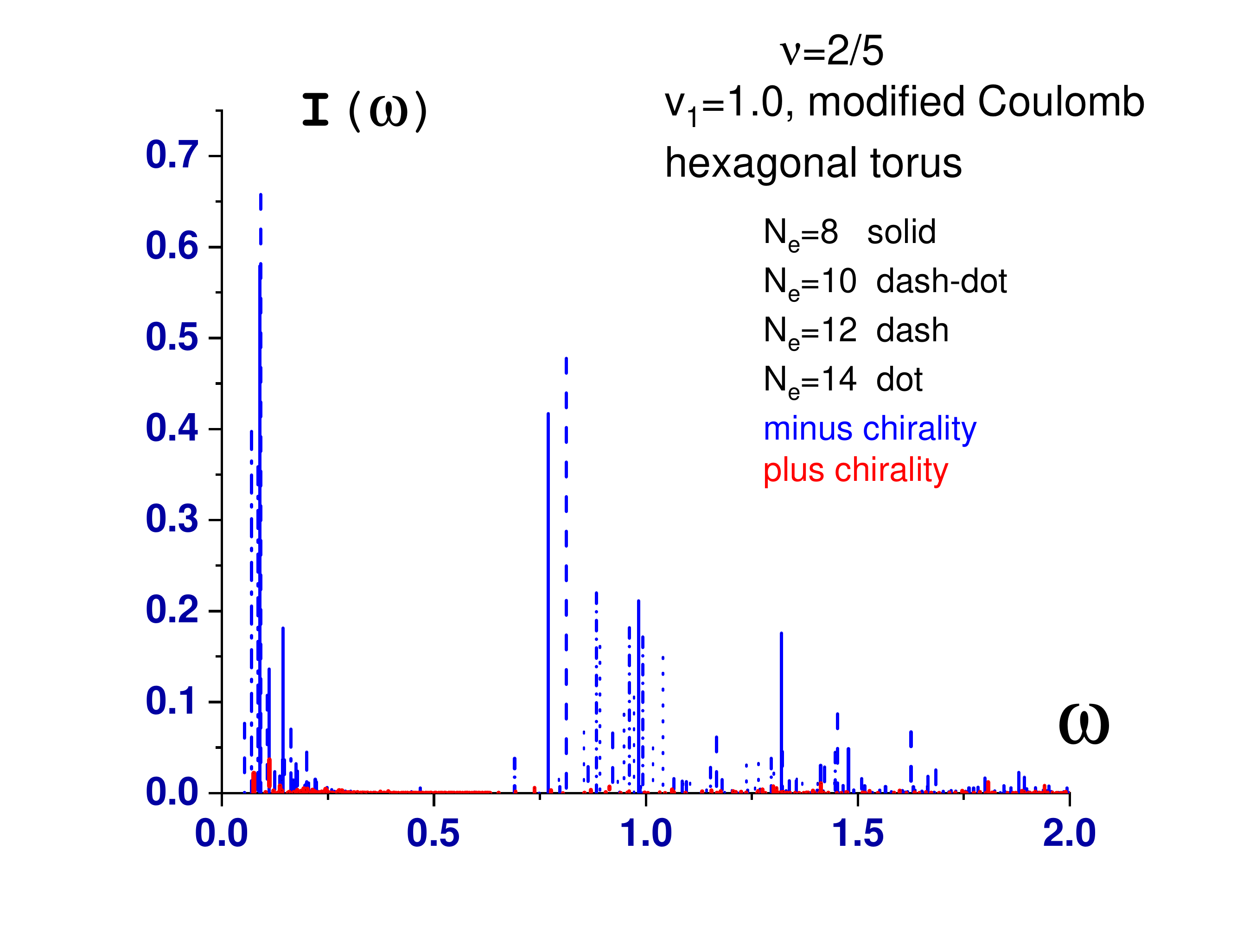}}
	\caption{The graviton spectral functions for the filling factor $\nu=2/5$ with the hard-core ($v_1=1$) and the modified Coulomb potentials. We compare these because they produce consistent values of $\bar{s}$ and $S_4$ in Fig \ref{fig:SR-2-5}. The spectrum on the left of 0.25 (in units of $e^2/(4\pi\epsilon_0\ell_B)$ is that of the modified Coulomb interactions and the spectrum to the right is for hard-core potential with $v_1=1$. The energies scales are different but the relative size of peaks appear to be mirrored in both cases. The modified Coulomb potential has the same scale and peak positions as the pure Coulomb potential in Fig \ref{fig:GS-2-5}. The graviton with the negative chirality dominates both spectral functions as expected.}
	\label{fig:supplemental2}   
\end{figure}

 We also include the graviton's spectral function for $\nu=2/5$, but for the hard-core ($v_1=1$) and the modified Coulomb potentials in Fig.~\ref{fig:supplemental2}.\vspace{-0.3cm} 

\end{document}